\documentclass[a4,10pt]{article}
\usepackage{amsmath,amsfonts,amssymb,graphicx,epsfig}
\usepackage{hyperref}
\usepackage{indentfirst}
 \date{}

\title{\textbf{A model for structural defects in nanomagnets}}

\author{F.A. Apolonio, W.A. Moura-Melo, F.P. Crisafulli, A.R. Pereira, and R.L. Silva
\\{\small Departamento de F\'{\i}sica,
Universidade Federal de Vi\c cosa, Vi\c cosa,  36570-000, Minas
Gerais, Brazil}}

\begin{document}
\maketitle
\begin{center}
\textbf{Abstract}
\end{center}

A model for describing structural pointlike defects in nanoscaled ferromagnetic materials is presented. Its details are explicitly developed whenever interacting with a vortex-like state comprised in a thin nanodisk. Among others, our model yields results for the vortex equilibrium position under the influence of several defects along with an external magnetic field in good qualitative agreement with experiments. We also discuss how such defects may affect the vortex motion, like its gyrotropic oscillation and dynamical polarization reversal.
Corresponding author: W.A. Moura-Melo \\
E-mail: winder@ufv.br\\
Tel.: +55-31-3899-3417.\\
Fax: +55-31-3899-2483.\\

\maketitle
\section{Introduction and Motivation}

\indent Nanomagnetism has become one of the most promising branches in Nanoscience and Nanotechnology. Several nanostructured magnets have been fabricated and a great deal of efforts has been employed to investigate their physical properties, specially those believed to have potentiality for applications, for example, in magnetoelectronic mechanisms for data recording and storage, ultra-precise magnetic sensors, and so forth\cite{nano1}.\\

\indent Among several systems, nanomagnets whose shapes comprise vortex-like magnetization, as stable remanent states, have received considerable attention. Actually, such a pattern minimizes the total energy (exchange + magnetostatic), for instance, in thin ferromagnetic nanodisks (with negligible anisotropy, like those made from Permalloy) \cite{Cowburn-etal-PRL83-1999,Ross-etal-JAP91-2002-etc,Shinjo-Science-Wacho,Guslienkovortice}. Several aspects of its structure and dynamics have been intensively investigated in the last years, like the interaction with litographically inserted defects \cite{Rahmtwoholes,Rahmtwoholes2,Afranio2005PRBeJAP,nossoJAP2007,KupperAPL2007,nossoJAP2008howhole,Cap-livro,JAPinterface2009}, core polarization reversal \cite{Nature444-2006,MertensPRL2006,YamadaNatureMat2007,reversaovelocidade}, and so forth. For example, the presence of one or more (large enough) defects considerably modify vortex profile and dynamics; generally, its core is captured by one of the defects after some time, so that a pinned vortex appears to be the final state \cite{Rahmtwoholes,Rahmtwoholes2,Afranio2005PRBeJAP,nossoJAP2007,KupperAPL2007,nossoJAP2008howhole,Cap-livro}. Core reversal, in turn, seems to be the one of the most important issues concerning vortex-like magnetization in nanomagnets, once the controlled reversion might open the doors for applications, like the utilization of the two polarization modes (up and down) as bits for recording and/or magnetic logic \cite{nano1,Nature444-2006,MertensPRL2006,YamadaNatureMat2007}. Nowadays, there are two distinct ways for achieving such a controlable reversal: applying suitable oscillating magnetic field pulses \cite{Nature444-2006}, or by means of spin polarized current \cite{MertensPRL2006,YamadaNatureMat2007}. It has been additionally proposed, and observed in simulations, that polarization switching can be induced by vortex-hole interaction\cite{nossoPRB2008} or even by an interface dividing a nanodisk into two halves \cite{JAPinterface2009}. Actually, it has recently elucidated that core reversal takes place, at least in pure and homogeneous samples, whenever the vortex core achieves a critical velocity \cite{reversaovelocidade}, $v_{\rm cr}\approx \, \gamma\, \sqrt{2A/ \mu_0}$ ($A$ is the exchange stifness while
$\gamma$ is the vortex gyromagnetic ratio; for typical Permalloy-made nanodisks $A\,\sim\, 10^{-11}\, {\rm J/m}$ and $\gamma\, \sim \,10^4\,{\rm Hz\,m/A}$, so that $v_{\rm cr}\,\sim\,10^2 \,{\rm m/s}$). Whether and how such a dynamical polarization switching is affected by both structural and litographically inserted defects remains untouched.\\

Even though these and many other issues should be investigated for a better comprehension of such configurations, it is noteworthy that even in the purest samples structural defects are present, generally randomly distributed throughout the material. In Ref. \cite{Compton-PRL2006}, authors estimate in $\sim 10^{11}/{\rm cm}^2$ their density, in usual Permalloy-made nanodisks. In this same work they also conclude that such defects may deeply modify vortex dynamics. In Ref. \cite{Hollinger-PRL95-2006}, it has been carried out an experiment  to analyze how those defects change vortex core path, as follows:  An external field is suitably applied in order to displace the vortex core; if the sample were pure, a plot of the vortex equilibrium position against this applied field would give a {\em perfect straight path}. However, authors found a very complicated path with pronounced roughness along it. Such a roughness, together with successive jumps, were attributed to (small) structural defects. Here, we would like to present a proposal for modeling such defects as they were point-like. Although simple in its conception, our theoretical model yields a counterpart for the above-mentioned path which is in good qualitative agreement with experiments. We also discuss on the two possible types of defects, acting as pinning (attractive) or scatter (repulsive) site, and the important role they can play in both gyrotropic oscillations and dynamical reversion of polarization by means of their interaction with the vortex core.\\

\section{The model and its basic results}

The basic ingredient in modeling such defects is taking them as they were pointlike. Therefore, each defect is formally represented by a quantity like $\lambda_i \delta^3(\vec{x}-\vec{x}_i)$, where $\lambda_i$ is a real parameter, so that $\lambda_i>0$ implies in a repulsive defect, while $\lambda_i<0$ means that it is attractive. Varying $\lambda_i$ we may control the strength of the interaction with the defect  located at $\vec{x}_i$. If we focus on isotropic materials, then the total energy of a magnet is the sum of the exchange and magnetostatic contributions, as below:
\begin{eqnarray}
E_{\rm net}= A \int_V \Big( \nabla\vec{m}\Big)^2\, dV + \frac{\mu_0 M_s}{2} \,\int_V \vec{m}(\vec{r}')\cdot \nabla \Phi_m(\vec{r}') d^3 \vec{r}'\,, \label{Enet}
\end{eqnarray}
where $V$ is the volume of the magnet with normalized magnetization $\vec{m}(\vec{r})=\vec{M}/M_s=(m_x,\, m_y,\, m_z)$. Above the scalar demagnetizing potential (as usually, we are supposing the absence of free electric currents throughout the magnet) is given formally in terms of the effective magnetic charges, say, their volumetric, $\rho_m=\nabla\cdot\vec{M}$ and superficial,$\sigma_m=\hat{n}\cdot\vec{M}$, contributions, like below:
\begin{equation}\label{Phimag}
\Phi_m(\vec{r})= -\frac{M_{s}}{4\pi}\int_{V'}\frac{\vec{\nabla'}\cdot\vec{m}(\vec{r}')}
{|\vec{r}-\vec{r}'|}dv'+\frac{M_{s}}{4\pi}\int_{S'}\frac{\hat{n'}\cdot\vec{m}(\vec{r}')}
{|\vec{r}-\vec{r}'|}da'\,.
\end{equation}

Of course, the explicitly expressions for both energies may be worked out if we fix the geometry of the magnet and its magnetization configuration. In addition, magnetostatic contribution, coming from the long-range dipolar-dipolar interaction, is generally not easy to be evaluated. Here, we shall examplify our model by explicitly working out the influences of those pointlike defects whenever interacting with a vortex-type magnetization in a thin ferromagnetic nanodisk. In addition, our description is performed within the so-called {\em rigid vortex regime}, which is strictly valid if the vortex profile is not appreciably deformed, say, if its core is not largely displaced from the nanodisk center. Although simplifying considerably our analysis, it should be remarked that the main physical aspects of the relevant interaction will be brought about. \\

Once the structural defects have been supposed to be pointlike, they will not appreciably alter the magnetostatic energy, since their volumes and areas, where effective magnetic charges should be computed, are strictly vanishing. This is further ensured provided that magnetization does not change abruptly around a given defect, which is the case for a vortex comprised in the magnet, as below. On the other hand, exchange term is modified, mainly at the vortex core, where such an energy is more concentrated. Actually, outside this core the defect-vortex interaction is approximately constant and much weaker than core-defect one. Thus, the important term in computing how the total vortex energy is modified whenever interacting with one or more defects is that concerning the core-defect interaction. Furthermore, once the vortex state will effectively interact only with those defects located on the nanodisk face, we essentially have a two-dimensional problem, so that $\lambda_i$ is effectively measured in area units. Typical values for $\lambda_i$ parameter should take into account experimental facts, reproducing them at least qualitatively and, from the theoretical point of view $\lambda_i$ must be very small compared to the typical area unity, brought about by the exchange length, in a such a way that $\lambda_i<< l^2_{\rm ex}$. Such a relation is fulfilled in our analysis: while $l^2_{\rm ex}\approx 10^{-17}\,{\rm m^2}$, usual values adopted here for $\lambda_i$ goes around $10^{-25}\,{\rm m^2}$ (see Figure 2 for further details). Then, while the defect-vortex interact, the parameter $\lambda_i$ show up indicating both its strength and if it is attractive or repulsive, as well. Eventually, such an interaction changes the exchange energy by $\Delta E_{\rm ex}=AL/2 \xi_i$, with:
\begin{equation}
\xi_i={\cal G}_i(\vec{r}-\vec{x}_i)\,\int \lambda_i \delta^2(\vec{x}'-\vec{x}_i) \Big(\nabla_{\vec{x}'} \vec{m}(\vec{x}')\Big)^2 d^2\vec{x}'\,. \label{xi}
\end{equation}  

\indent Here ${\cal G}_i$ is a function which accounts for how the defect located at $\vec{x}_i$ interacts with the vortex core. Although its exact profile is difficult to be determined in practice, a gaussian function seems to be a reasonable choice, ${\cal G}_i(\vec{r}-\vec{x}_i)=e^{-\alpha (\vec{r}-\vec{x}_i)^2}$. [Other functions could be used to simulate such an interaction; disregarding its precise form, a free parameter, like $\alpha$, should be introduced.] Its presence here essentially ensures that the vortex-defect interaction is smooth, starting effectively when the core border meets the defect, raising until a maximum, where the vortex center strikes the defect, then lowering. In addition, it plays also the role of controlling the interaction range to reasonable values, around the vortex core radius, $l_0$, in our case (see Figure 1). For Permalloy-made samples \cite{Usov-KurkinaJMMM}, $l_0\approx 0.68 l_{\rm ex}(L/l_{\rm ex})^{1/3} \approx 8-15 \, {\rm nm}$, so that the range of this interaction $\delta=2l_0/e\,\sim\,10 \, {\rm nm}$, in agreement with experimental findings \cite{Compton-PRL2006,Hollinger-PRL95-2006}. Above, we have assumed disks with thickness $L\sim 10^1\, -\, 10^2\, {\rm nm}$ and $l_{\rm ex}=\sqrt{ 2A/\mu_0 M^2_s}\sim\, 5\,-\, 6 {\rm nm}$, once the saturation magnetization reads $M_s\approx 8\,\times\,10^{5}\, {\rm A/m}$, for this compound.\\

In eq. (\ref{xi}), $\vec{m}=\vec{M}/M_s=(\sin\theta\,\cos\phi; \sin\theta\,\sin\phi; \cos\theta)$ is the classical unit magnetization vector. In order to describe a vortex-type state with non-vanishing polarization, we must have $\phi\equiv \phi_v=Q\tan^{-1}(y/x) + q \,\pi/2$, where $Q$ is the vorticity of the solution (topological winding number; $Q$ is an integer) while $q=\pm 1$ is the vortex chirality ($q=-1$ or $q=+1$ for clockwise or counter-clockwise circulation of the magnetization). On the other hand, $\theta$-variable describing the vortex state is exactly known only asymptotically: $\cos\theta_v\to \pm1$ at the vortex center, $\vec{r}=0$, diminishing as one goes outside the core, where it vanishes. Among several suitable trial functions, and without loss of generality, we may take $\theta_v$ to be:
\begin{equation}
\theta_v=\left\{\begin{array}{l} 
                     \pi/2\;\; {\rm if} \;\; r>l_0 \\
                   p\,\cos^{-1}\left(1-\frac{r^2}{l^2_0}\right)^n \;\; {\rm if} \;\;  r\le l_0\;,                                           
\end{array} \right.
\end{equation}
where $p=\pm1$ is the vortex polarization ($+$ if the moment at the center points {\em up} or $-$ if {\em down}). The parameter $n$ may be choosen for adjusting the vortex core profile (see Ref.\cite{Bahiana}, for additional details). For simplicity, we shall take $n=1$ in all of our calculations hereafter. Taking all these ingredients to eq. (\ref{xi}), we readily obtain: 
\begin{equation}
\xi_i= \frac{(2l^2_0 -r^2)^2+ 4l^4_0}{l^4_0(2l^2_0 -r^2)}\, \lambda_i {\cal G}_i(\vec{r}-\vec{x}_i) \,.\label{xifinal}
\end{equation}
Therefore, the normalized energy density, $u_{\rm net}=E_{\rm net}/4\pi \mu_0 M^2_0 V$ of the nanomagnet comprising a vortex interacting with a fixed defect at $\vec{x}_i$, within the rigid vortex model, will read as below (the generalization for $N$ such non-interacting defects may be readily get):
\begin{equation}
u_{\rm net}= \frac{l^2_{\rm ex}}{R^2}\,\ln(1-s^2)\, [1-\xi_i] +2\pi F_1(L/R)\, s^2 - h_{\rm ext} \, s\,, \label{unet}
\end{equation}
where  $R$ and $L$ are the radius and thickness of the nanodisk. Parameter $s$ measures the relative displacement of the vortex core from the geometrical center of the nanodisk, $\vec{s}\equiv \vec{r}/R$ (rigid vortex is ensured for small $s$; reasonale results still emerge at intermediary values, $s \sim 0.2-0.3$). Zeeman effect is accounted by the normalized field $h_{\rm ext}=H_{\rm ext}/M_s$. In turn, $F_1$ is an auxiliary function, defined by: $F_1(z)=\int^\infty_0 J^2_1(t)\,(e^{-zt}+zt -1) dt/ zt^2$. Note also that the normalized energy is computed relative to the origin, $\vec{s}=0$, so that it does not give the precise value of the interaction between the vortex core and a defect located there. Here, such a value is normalized to zero, but it can be obtained explicitly by computing the exchange energy associated to the vortex interacting with a defect $E_{\rm ex-int}=E_{\rm ex}-E_{\rm int}=E_{\rm ex} (1-\xi_i)$. For our purposes, $u_{\rm net}$ is the relevant quantity. Now, expanding the exchange term in eq. (\ref{unet}) around $s=0$ and retaining only those contributions up to $s^2$, we finally obtain that the equilibrium position of the vortex core, subject to an external field and interacting with a defect, reads:
\begin{equation}
s_{\rm eq}=\frac{h_{\rm ext}}{4\pi F_1(L/R) -(1-\xi_i)(l_{\rm ex}/R)^2}\,. \label{seq}
\end{equation}
Clearly, expressions (\ref{unet}-\ref{seq}) recover their usual forms \cite{Guslienkovortice} as long as we remove the defect, $\xi_i=0$. \\

\section{Further results and discussion}

Now, let us consider a number of defects, for instance along a given line, and apply a suitable field for moving the vortex core towards each defect. This will enables us to show how our model describes vortex-defect interaction more clearly, making possible a comparison with available experimental findings. For that, let us place several pointlike fixed defects, with different interaction parameters $\lambda$, along $x$-axis. An external field is then applied to move vortex core along this line, so that it interacts with these defects while moving. The resulting plot of its equilibrium position against the applied field is shown in Figure 2. Note that, by virtue of the interaction with the defects, the vortex core trajectory is no longer a straight line, which is the path for a pure sample, with $\xi_i=0$ in eq. (\ref{seq}). Besides of being quite complicated it should be noted that this actual trajectory presents a remarkable roughness, mainly around each defect position. We also note two distinct sorts of jumps in this plot: i) abrupt changes in $s_{\rm eq}$, at practically the same value of $h_{\rm ext}$, what says us that the core is interacting with a repulsive defect, $\lambda>0$ (indicated by black arrows in Fig. 2); ii) conversely, in other places $s_{\rm eq}$ is kept practically unaltered while $h_{\rm ext}$ changes considerably, evidencing the interaction with an attractive defect, $\lambda<0$ (as blue arrows indicate in Fig. 2). Comparing these with the experimental results provided in Ref.\cite{Hollinger-PRL95-2006}, mainly with Figure 5 and related text from this article, a good qualitative agreement between them can be realized.\\

Once the vortex core position is changed whenever interacting with scatter or pinning defects throughout the sample, we may wonder whether and how these structural defects could modify the gyrotropic motion and its associated frequency. Qualitatively, it is expected that its traced path will be much more complicated than that smooth ones in pure samples. The roughness along its actual path will resembles us a Brownian-type motion, enlarging the distance followed by the vortex to complete a revolution. If the concentration of pinning and scatter sites is similar each other, a net compensation is expected so that the gyrotropic frequency is expected to remain practically unaffected, provided that the scattering does not enlarge the trajectory of the core. However, if we have much more, say, pinning sites, a net retardation in the vortex motion should be verified causing a decreasing in the frequency. Such an effect could be estimated qualitatively, for instance, if we consider the Landau-Lifshitz equation:
\begin{equation}
\vec{G}\times \vec{v}=k \vec{r} + m_v \vec{a}\,,\nonumber
\end{equation}
with the reasonable assumption that the vortex mass is very small  \cite{Cap-livro}, $m_v \approx 0$. In this equation, $\vec{G}=-(M_s/\gamma)\int_V \Big[\nabla(\cos\theta_v) \times\nabla\phi_v \Big] \, dV$ is the gyrotropic vector of the vortex, with $\theta_v$ and $\phi_v$ being the vortex state functions, previously given. In the simplest case of a vortex in a circular nanodisk it may be evaluated to give $\vec{G}=(2\pi M_s L/\gamma) pq\hat{z}$, evidencing its topological nature. Parameters $\vec{r}$, $\vec{v}$ and $\vec{a}$ are the position, velocity and acceleration of the vortex core, while $k$ plays the role of a spring-like constant, derived from the net energy of the magnet as usual, $\vec{F}=-k\vec{r}=-\nabla E_{\rm net}$. In this case, supposing small vortex core displacement, $s<<1$, $k$ is explicitly obtained to be:
$$
k= \pi L M^2_s \Big[ 4\pi F_1(L/R) -(1-\xi_i)\frac{l^2_{\rm ex}}{R^2}\Big]\,.\nonumber
$$
Once $k$ is changed by the $\xi$-factor, we clearly see that the structural defects are expected to modify the gyrotropic motion of the vortex core. However, we cannot proceed further with our analysis because we can no longer assume that the vortex core perform harmonic oscillations around its equilibrium position (as is usually done), once deviations may be appreciable whenever defects are present, making our analytical approach efficientless; by virtue of that, numerical simulation appears to be a more suitable way for studying how these defects affect the gyrotropic frequency \cite{workinprogress}. Moreover, they can also change the situation for core reversal take place. Actually, even assuming that defects does not change the critical velocity (in pure samples, it only depends on the exchange stiffness, $v_{\rm cr}\approx \gamma \sqrt{2A/\mu_0}$ \cite{reversaovelocidade}) they may considerably change the scenario of polarization switching: For instance, if vortex core is moving in a region with larger concentration of attractive defects, the successive pinnings will deaccelerate its core, dissipating kinetic energy as spin waves and/or heat, preventing its immediate achieving of the critical velocity. Similarly to the case of gyrotropic motion, a precise answer demands further investigation, namely, by means of numerical simulations.\\

\section{Conclusions and Prospects}

A model for structural defects in nanoscaled ferromagnetic systems is presented. Once such defects are described as pointlike objetcs their main modification appears in the exchange contribution to the total energy. As an example, we have worked out the case of a vortex-type pattern comprised in a thin nanodisk. An expression for its equilibrium position whenever subject to both, an external field and  defects, is analytically obtained, immediately recovering its usual counterpart if the defects  influence is removed. A good qualitative agreement between experimental and our theoretical results, concerning how structural defects modify the vortex motion, is clearly realized.\\

Among other prospects for future investigation, we may quote the refinement of some ingredients entering this model (defect-magnetization interaction form, strength of the interaction, and so forth), for instance, to explicitly work out how such defects affect other magnetization configuration and/or sample geometries. We also intend to reconsider this model, but for studying magnetocristaline anisotropy. In this case, we expect that instead of  $\lambda_i$ we should consider a vector function, $\vec{\Lambda}(\vec{x})$, which should account for the preference of the magnetization in pointing in a given direction at each point or region inside the nanomagnet. Such a study could perhaps shed some extra light on the pronounced asymmetry in the gyrotropic frequency between up and down polarizations, as observed in the experiments of Ref.\cite{Hoffmann-PRB2007}, and very recently addressed in the work of Ref. \cite{asymmetry-updown}.\\

\centerline{\large Acknowlegments\\}
\vskip .5cm
\indent The authors are grateful to CNPq and FAPEMIG for financial support.
\newpage 
\centerline{\bf Figure Captions\\}

Figure 1:  Illustrates how the defect-vortex interaction takes place: tiny at the core borders, where exchange energy is small, raising to its maximum at the vortex center, where exchange density gets its highest value. At $r=0$ the interaction is maximum whose strength reads $4\lambda_i/l^2_0$. Its effective range $\delta$ is around $10\,{\rm nm}$, in accordance with recent experiments; its (Gaussian) shape may be controlled by varying $\alpha$-parameter (we have taken $\alpha^{-1}=2.5\,\times\, 10^4$ in order to set the interaction range $\sim 10\,{\rm nm}$, according to experiments).\\

Figure 2 (Color online): How the vortex equilibrium position versus external field is modified by virtue of the interaction with structural defects (if the sample were pure, a straight line would be the case). A comparison with experimental findings, as reported in Ref. \cite{Hollinger-PRL95-2006}, shows a good qualitative agreement. Namely, note the interaction with pinning and repulsive defects, indicated by blue and black arrows, respectively. [Typical values used for $|\lambda_i|$ read in the range $\sim 10^{-25}\,-\, 10^{-26}\, {\rm m^{2}}$. Note that such values are much smaller than $l^2_{\rm ex}\sim 10^{-17}\, {\rm m^2}$, the elementary exchange area, showing the smallness of our described defects, as initially supposed].

\begin{figure}[!h]
\centering 
{\includegraphics[width=6cm,angle=90]{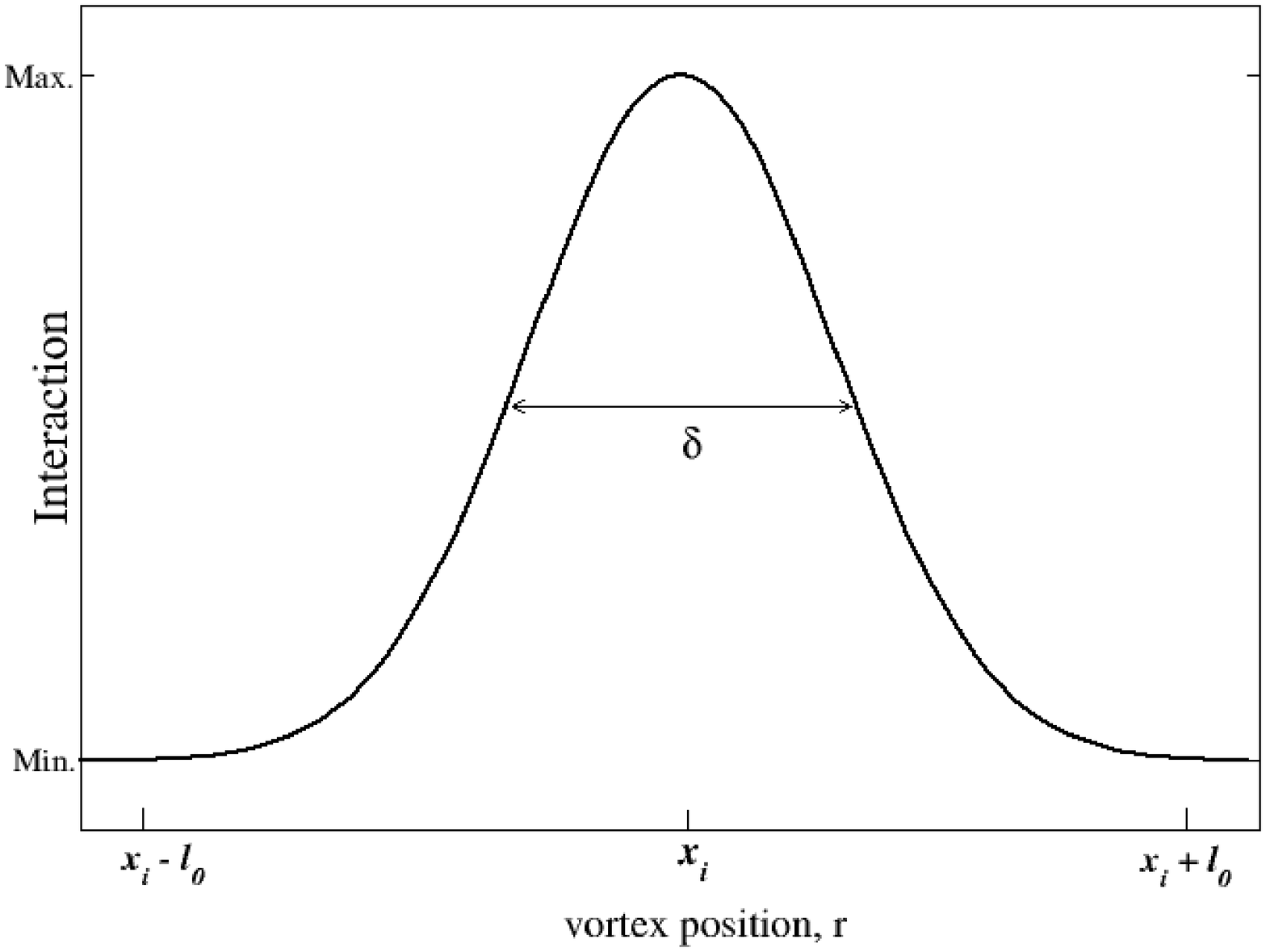}}
\caption{{\protect\small}} \label{fig1}
\end{figure}

\begin{figure}[!h]
\centering 
{\includegraphics[width=6cm,angle=90]{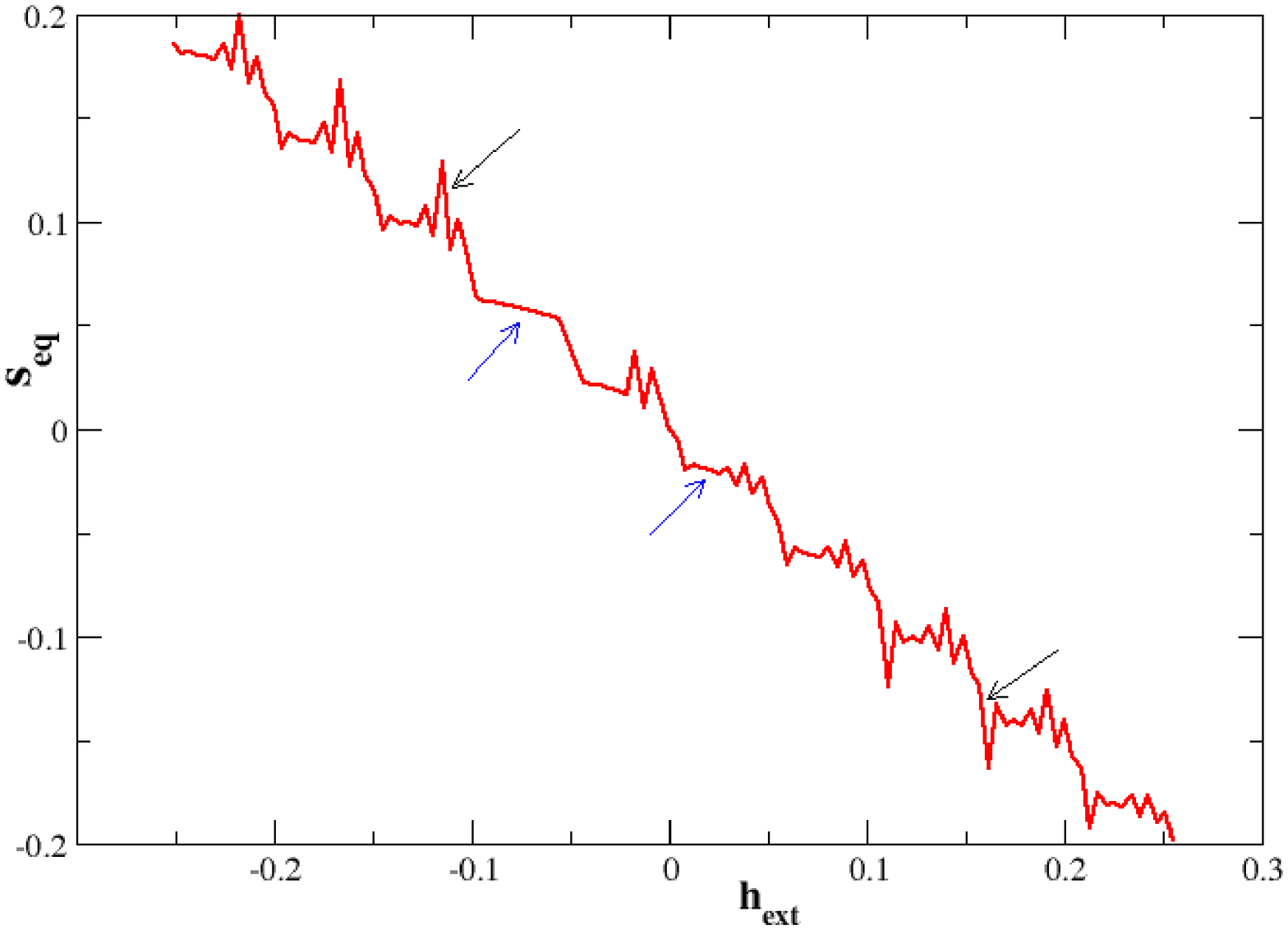}}
\caption{{\protect\small  }} \label{fig2}
\end{figure}

\end{document}